\documentclass[prb, twocolumn,showpacs]{revtex4}
\begin{document}

%\title{Ballistic Electron Spectroscopy of Individual Buried Molecules}

\title{Can Ballistic Electrons Probe the Electronic Spectra of Individual Buried Molecules?}

\author{George Kirczenow}

\affiliation{Department of Physics, Simon Fraser
University, Burnaby, British Columbia, Canada V5A 1S6}

\date{\today}

\begin{abstract}
A theoretical study is presented of the ballistic electron
emission spectra (BEES) of individual insulating and
conducting organic molecules chemisorbed on a silicon
substrate and buried under a thin gold film. It is
predicted that ballistic electrons injected into the gold
film from a  scanning tunneling microscope tip should be
transmitted so weakly to the silicon substrate by alkane
molecules of moderate length (decane, hexane) and their
thiolates that individual buried molecules of this type
will be difficult to detect in BEES experiments. However,
resonant transmission by molecules containing unsaturated
C-C bonds or aromatic rings is predicted to be strong
enough for BEES spectra of individual buried molecules of
these types types to be measured. 
Calculated BEES spectra of molecules of both types are
presented and the effects of some simple
interstitial and substitutional gold defects that
may occur in molecular films are also briefly
discussed.       

\end{abstract}
% 81.07.Nb Molecular nanostructures
% 81.07.Pr Organic-inorganic hybrid nanostructures
% 73.23.Ad Ballistic transport 
% 73.63.-b Electronic transport in nanoscale materials 
% and structures

\maketitle

\section{Introduction}

During the last decade there has been growing interest
in molecular electronics, stimulated largely by the
experimental realization of molecular
wires\cite{Bumm,Reed1,Datta}, systems in which a single
organic molecule or a few molecules carry an electric
current between a  pair of metal contacts. Hybrid
molecule/silicon nanoelectronic devices are another
intriguing possibility and research that may  lead to
their creation is also  being
pursued.\cite{Lopinski,Cho,Hofer,Hersam,Tong,Rakshit,Wang,Nature,silicon}
For many potential  nanoelectronic applications it
will be necessary to sandwich a molecular layer
between metal and/or semiconductor layers that act
as contacts  and also protect the molecular layer
from the environment. However, this makes the
molecules inaccessible to direct nanoscopic
experimental  probes such as the scanning tunneling
microscope (STM). Thus the structures and electronic
properties of such buried molecular layers are, to a
large degree, still the subject of conjecture.

Ballistic electron emission microscopy (BEEM) is a
technique that can probe the nanoscale spatial
structure and electronic spectra of buried
interfaces.\cite{Bell-Kaiser,BEEMreview} In BEEM
experiments a bias voltage is applied between an
STM tip and a thin metal overlayer covering a
buried substrate, usually a semiconductor, which
is kept at almost the same electrochemical
potential as the metal overlayer.\cite{meV}
Electrons are injected from the STM tip into the
metal layer and travel ballistically through the
metal. Some of them are transmitted through the
metal/semiconductor interface into the
semiconductor substrate.The electric current
transmitted through the interface is measured as a
function of both the lateral position of the STM
tip and the bias voltage applied between the tip
and metal overlayer. This allows the imaging of
defect structures at the interface and
spatially-resolved spectroscopy of their
electronic structure, with nanometer resolution.
Recently BEEM experiments have been initiated on 
systems in which self-assembled molecular
monolayers are present between the metal overlayer
and semiconductor, specifically, alkane
dithiolates between GaAs and
gold.\cite{KavanaghGaAs} Since in such BEEM
experiments a significant bias is not applied
across the molecular layer, BEEM spectroscopy
(BEES) of buried molecules (unlike the usual
spectroscopies based on measurements of
current-voltage  characteristics of molecular
diodes\cite{Bumm,Reed1,Datta}) is not subject to
complications due to bias-induced charging effects
and electric fields in or near the molecular
layer. Thus in addition to being a unique probe of
buried structures with nanometer spatial
resolution, BEEM can provide important
spectroscopic information relevant to molecular
electronics that is complementary to that obtained
from measurements of diode current-voltage
characteristics.    However, because of the absence to
date of any appropriate theory, it has been
difficult to interpret the results of BEEM
experiments for metal/molecule/semiconductor
systems unambiguously: For example, in the recent
experimental study\cite{KavanaghGaAs} a BEEM
signal could only be detected in a few small,
isolated patches of the sample and it has been
unclear whether this should be interpreted as
evidence that most of the sample was occupied by
an insulating alkane-dithiolate self-assembled monolayer
(SAM) that did
not transmit a detectable BEEM signal and that the
observed signal was coming from occasional patches
of defects in this SAM, or that to the contrary
the detected BEEM signal was being transmitted
through small patches of alkane-dithiolate SAM
surrounded by much larger regions of a more
strongly insulating material (such as
alkane-dithiol multilayers) between the
semiconductor and the gold overlayer. 

In this article a theoretical study is presented
of ballistic electron emission
microscopy/spectroscopy of single organic
molecules between a thin gold overlayer and a
silicon substrate. It is predicted that alkane
molecules of moderate size (hexane and decane are
studied) attached to the Si substrate by a C-Si
bond (and whether thiol-bonded or not chemically
bonded to the gold overlayer) should transmit the
ballistic electrons so weakly that such single
molecules are expected to be below the threshold
of detection for conventional BEEM equipment.
However, related single molecules containing
unsaturated C-C bonds or aromatic rings are
predicted to exhibit sufficiently strong
transmission resonances that their observation in
BEEM experiments should be feasible. Thus it is
predicted that $single$ molecules with unsaturated
C-C bonds or aromatic rings embedded in an alkane
or alkane thiolate SAM between a silicon substrate
and gold overlayer should be detectable in BEEM
experiments with present day equipment. The
calculated BEEM spectra of the molecules are also
presented and the effects of some simple
interstitial and substitutional gold defects that
may occur in molecular films are also briefly
discussed. 

This paper is organized as follows: The
theoretical model of the structures and electronic
structures of the systems studied and the approach
used to calculate the BEEM currents in these
systems are described in Section \ref{Model}. The
results obtained are presented together with their
interpretation and implications for experiment in
Section \ref{Results}. The main conclusions are
summarized in Section
\ref{conclusions}.   

\section{Model and Theoretical Approach}
\label{Model}

\subsection{Extended Molecule}
\label{Extended_Molecule}  

As in other transport calculations in molecular 
electronics,\cite{Datta,Rakshit,silicon,metalcontacttheory1,Emberly,metalcontacttheory2,Kushmeric,metalcontacttheory3,Dalgleish}
the theoretical model studied is based on an
$extended$ $molecule$ that includes the molecule
itself as well as clusters of nearby atoms
belonging to the electrodes that transmit
electrons to and from the molecule. In this way
the chemical bonding between the molecule and the
electrodes is taken into account as well as the
electronic structures of the molecule and the
electrodes. In the present systems there are three
electrodes: the STM tip, the metal layer between
the tip and the molecule, and the silicon
substrate. An atomic cluster from each of these is
included in the extended molecule, as is shown for
an ethylmethylbenzene molecule in Fig.\ref{Fig_1}:
A cluster of 10 Au atoms represents the
(111)-oriented monoatomic Au tip. The Au film
between the tip and molecule is represented by a
591 Au atom cluster that is approximately
cylindrical and presents (111)-oriented facets to
the STM tip and the molecule. A hemispherical 390
Si atom cluster with the dangling Si bonds
passivated by H atoms and its flat (111) surface
facing the molecule represents the Si substrate.
The molecules studied  here bond to the Si
substrate through a single covalent C-Si bond. For
those molecules that are thiol-bonded to the gold
film, the sulfur atom  is positioned $\sim$ 
{2.2\AA} below a
hollow site between three gold surface atoms.
Density functional theory was used to estimate the
atomic geometries of the molecules and of their
bonding to the semiconductor substrate and metal
overlayer.\cite{Gausstip} However in cases where
the end of the molecule adjacent to the gold layer
terminates in a CH$_3$ group (which does not bond
to the gold), the distance between the gold
surface plane and the closest H atom of the
molecule is somewhat arbitrary, and was assumed to
be 3\AA.

\subsection{Electronic Structure}
\label{Electronic_Structure}

Most theoretical work on electronic transport in
molecular  wires with metal contacts has been
based on  semi-empirical tight-binding models or
Kohn-Sham density functional calculations of the
electronic
structure.\cite{Datta,metalcontacttheory1,Emberly,metalcontacttheory2,Kushmeric,metalcontacttheory3,Dalgleish}
It was found in the present study that, in order
to realistically model ballistic transmission of
electrons through the gold film between the STM
tip and molecule  in BEEM experiments, several
hundred Au atoms need to be included in the
cluster that represents the gold film in extended
molecule. This makes the density functional
approach impractical for the transport problem. 
Furthermore previous theoretical work modeling the
current-voltage characteristics of molecules on
silicon found Kohn-Sham density functional
calculations to be unsuitable for treating the
electronic structure of the silicon
substrate,\cite{Rakshit} and to yield results that
disagree qualitatively with experimental STM
data.\cite{silicon} However, transport
calculations based on semi-empirical tight-binding
models  can be performed for systems with large
numbers of atoms and have successfully explained 
experimental current-voltage characteristics of
molecules connecting gold
electrodes\cite{Datta,Emberly,Kushmeric} and of
molecular wires on silicon
substrates.\cite{silicon} Thus a hybrid
tight-binding model of the electronic structure
that combines  parameters derived from
semi-empirical considerations of quantum chemistry
and from {\it ab initio} band structure
calculations is adopted here. As is explained
below, a number of parameters that enter the model
are not known accurately, and in order to proceed
further it is necessary to make plausible but not
rigorous approximations and assumptions regarding
their values. The sensitivity of the present
findings to the values of these parameters was
explored in the course of this work and it was
found that the {\em qualitative} conclusions
reported here are robust to reasonable variations
in parameter values. The reader is referred to
previous papers\cite{silicon,Dalgleish,nanotube}
for extended discussions of the reliability of
such techniques. 

The semi-empirical extended H{\"u}ckel model of
quantum chemistry\cite{EH,YAeHMOP} is used to describe
the electronic structure of the molecule and coupling
between the molecule and silicon substrate. However in
order to model the electronic structure of the silicon
substrate satisfactorily, the extended H{\"u}ckel
model requires modification and this is done here as
in Ref.\onlinecite{silicon}: Extended H{\"u}ckel
theory describes molecular systems in terms of a small
set of    Slater-type atomic orbitals $\{
|\phi_i\rangle \}$, their overlaps
$S_{ij} =
\langle\phi_i | \phi_j\rangle$ and a Hamiltonian matrix
$H_{ij} =
\langle\phi_i |H| \phi_j\rangle$. The diagonal
Hamiltonian elements $H_{ii} = \epsilon_i$  are the
atomic orbital ionization energies. Non-diagonal
elements are approximated by 
\begin{equation}
H_{ij} = K
S_{ij}(\epsilon_i + \epsilon_j)/2
\label{Hij}
\end{equation}
where $K$ is a phenomenological parameter usually
chosen to be 1.75 for consistency with experimental
molecular electronic structure data. However, in the
present work, as is discussed in detail in
Ref.\onlinecite{silicon}, for Si atoms the single
parameter $K$ is replaced by a set of parameters
$K_{ij}$ whose values are chosen so that the modified
extended H{\"u}ckel model obtained in this way yields
an accurate description of the band structure of 
crystalline silicon. The electronic structures of the
gold film and STM tip are also described by a
tight-binding model with a non-orthogonal basis but in
this case the tight-binding parameters $H_{ij}$ and
$S_{ij}$ from Ref. \onlinecite{Papaconstantopoulos}
that are based on fits to
$ab$ $initio$ calculations of the electronic structure
of gold are used. The atomic orbital energies $H_{ii}$
of Ref. \onlinecite{Papaconstantopoulos} for gold are
defined up to an arbitrary additive constant which is
chosen in the present work so as to yield reasonable
values of the energy offsets
$\Delta_{M,Au}$ between the HOMO levels of the
molecules $M$ studied and the Fermi level of the gold
film. The values of the offsets $\Delta_{M,Au}$ are
not known accurately at present either theoretically
or experimentally, their determination being an
important open problem of molecular electronics. For
the present purpose they are estimated as the
differences between the work function of Au and the
HOMO energies of the respective isolated molecules
obtained from density functional theory, an
approximation that has been used successfully to align
the Fermi energy of gold with the HOMO of
benzenedithiolate,\cite{Derosa2001} and has also been
applied to a variety of other metal-molecule
junctions.\cite{Dalgleish} The effects of the
potential difference between the STM tip and the gold 
film on the electronic structure of the system are
included in the present model  by offsetting the
atomic orbital energies $H_{ii}$ for gold STM tip
upwards in energy from those of the gold film by an
amount
$|eV|$ where $V$ is bias voltage between the tip and
Au film. The value of the energy offset
$\Delta_{Si,Au}$ of the silicon conduction band edge
(at the surface of the silicon substrate)  above the
Fermi level of the gold film is another important
parameter that is not known accurately for
gold/molecule/silicon heterojunctions. Here it is
assumed, for simplicity, that $\Delta_{Si,Au}=0.84eV$,
a typical experimental value of the Schottky barrier
measured in BEEM experiments  on Au/Si(111)
interfaces;\cite{offset} the Si orbital energies
$H_{ii} = \epsilon_i$ are adjusted (all equally)  so
as to place the LUMO level of the Si cluster of the
extended   molecule $0.84eV$ above the Fermi level of
the gold film. Finally, the Hamiltonian matrix
elements between the STM tip and gold film and between
the gold film and molecule are estimated from Eq.(\ref
{Hij}) with $K= 1.75$. Because of the
non-orthogonality of the basis states employed in the
present model, the above adjustments of the diagonal
matrix elements $H_{ii}$ of the tight binding
Hamiltonian require\cite{silicon,shift} corresponding
adjustments
$\Delta H_{ij}$ of the non-diagonal elements
$H_{ij}$, which were taken to be  
\begin{equation}
\Delta H_{ij} =
S_{ij}(E_i + E_j)/2
\label{gauge}
\end{equation}
 where $E_k$ is the shift of the diagonal element
$H_{kk}$.

\subsection{Transport}
\label{Transport}

In a BEEM experiment the electron flux that passes
from the STM tip to the gold film subsequently
separates: A small part of it (that constitutes the
BEEM current
$I_{\mathrm{BEEM}}$) continues through the molecule
to the Si substrate and on to the electron drain
electrode, while most of it (referred to henceforth
as the film current
$I_{\mathrm{FILM}}$) proceeds laterally through the
gold film directly to the drain electrode. The total
current $I_{\mathrm{tip}}$ passing through the STM
tip is thus given by
$I_{\mathrm{tip}}=I_{\mathrm{BEEM}}+I_{\mathrm{FILM}}$.
 
The present calculations of $I_{\mathrm{BEEM}}$ and
$I_{\mathrm{FILM}}$ are based on
Landauer-B{\"u}ttiker theory.\cite{book}  For BEEM
experiments the gold film, the Si substrate and the
electron drain can all be considered to be at the
same electrochemical potential.\cite{meV} Also, in
practice, electron transmission from the source to
drain via the gold film, molecule and silicon
substrate, and directly via the gold film are
mutually incoherent processes. Therefore it follows
from Landauer-B{\"u}ttiker theory that the currents
$I_{\mathrm{BEEM}}$ and $I_{\mathrm{FILM}}$ at a
given value of the applied source-drain bias voltage
$V$ are related to the multichannel electron
transmission probabilities
$T_{x}(E,V)$ at energy $E$ from the source electrode
to the drain via $x$ where $x$ can be the BEEM path
through the gold film, molecule and Si substrate ($x
= {\mathrm{BEEM}}$), or the direct path
 through the gold film to the drain ($x =
{\mathrm{FILM}}$), according to
\begin{equation}
I_x(V) = -\frac{2e}{h} \int_{-\infty}^{\infty}
dE\:T_{x}(E,V)\left( f(E,\mu_{s}) - f(E,\mu_{d})\right)
\label{Landauer}
\end{equation}
where $f(E,\mu_{i}) ={1}/{(\exp[(E-\mu_{i})/kT] + 1)}$
and $\mu_{i}$ is the electrochemical potential of the
source ($i=s$) or drain ($i=d$) electrode.\cite{temp}

The source and drain electrodes are modelled here as
arrays of ideal single-channel leads: One such drain
lead is coupled to the 1$s$ orbital of each of the
hydrogen atoms that passivate the dangling bonds
terminating the Si cluster of the extended molecule,
and one is coupled to each $p$ valence orbital of the
surface atoms of the Si cluster (except for those of
the H and Si atoms in the vicinity of the Si atom to
which the molecule bonds), a total of 675 ideal
leads. An ideal drain lead is also coupled to each
$s$, $p$ and $d$ valence orbital of the surface atoms
of the Au cluster representing the the Au film,
except for those  Au atoms in the immediate vicinity
of the molecule and STM tip, totalling 2709 leads.
Similarly an ideal source lead is coupled to each
valence $s$, $p$  and $d$ orbital of the surface
atoms of the cluster representing the STM tip, except
for the terminal atom of the tip, a total of 81
leads.         Each ideal lead $k$ was modelled as a
semi-infinite tight-binding chain with a single
orbital per site, a site energy
$\alpha_k$ and nearest neighbor hopping matrix element
$\beta$. $\alpha_k$ was chosen to be equal to the
energy (including the shift, if any, due to the
applied bias) of the hydrogen, silicon or gold
orbital to which lead $k$ was coupled. The hopping
matrix elements $\beta$ were all taken to have a
magnitude of 6 eV, sufficiently large that the
eigenmodes of all of the ideal leads in the energy
window between $\mu_{s}$ and
$\mu_{d}$ in eq. (\ref{Landauer}) be composed of
counter-propagating (left and right moving) waves. The
coupling matrix elements
$W_k$ between the ideal leads and their respective
hydrogen, silicon and gold orbitals were also set to
$\beta$. 

As well as mimicking macroscopic electrodes by
transmitting an ample electron flux to and from the
system, the above large numbers of ideal source and
drain leads have a similar effect to
phase-randomizing B{\"u}ttiker
probes\cite{Buttikerprobes} in minimizing the
influence of dimensional resonances due to the finite
sizes of the gold and silicon clusters employed in
the model; it was verified that the ideal leads
described above meet these requirements by comparing
the results obtained in test calculations with
differing numbers of ideal leads, differing values of
the lead parameters, and differing numbers of atoms
in the Au and Si clusters of the extended molecules.
Calculations of the currents $I_{\mathrm{BEEM}}$  and
$I_{\mathrm{FILM}}$ for cylindrical Au atom clusters
with  the same height (1.4nm) as that of the 591 Au
atom cluster  in Fig.\ref{Fig_1} but with differing
diameters confirmed that the diameter of this
cluster (approximately 2.5nm) was sufficient for
these currents to have converged with increasing
cluster diameter, i.e., that electron transport
through this  finite cluster adequately represents
ballistic propagation of electrons through a thin but
laterally extended gold film.

To calculate $T_{x}(E,V)$ and hence evaluate
eq.(\ref{Landauer}), the transformation to the
alternate Hilbert space described in Refs.
\onlinecite{orthog1} and
\cite{orthog2} was made, mapping the non-orthogonal
basis of atomic orbitals to an orthogonal basis. The
Lippmann-Schwinger equation 
\begin{equation}
|\Psi_{k}\rangle = |\Phi_{o,k}\rangle + G_{o}(E) W
|\Psi_{k}\rangle
\label{eq:LSch}
\end{equation}
describing electron scattering between the source and
drain leads was solved numerically for
$|\Psi_{k}\rangle$ in the alternate Hilbert
space.\cite{orthog1,orthog2} In eq.(\ref{eq:LSch}) 
$G_{o}(E)$ is the Green's function for the decoupled
system (i.e., with the coupling  between the ideal
leads and the extended molecule switched off),
$|\Phi_{o,k}\rangle$ is the eigenstate of the
decoupled ideal source lead $k$ with energy
$E$ and
$|\Psi_{k}\rangle$ is the corresponding scattering
eigenstate of the complete system with the coupling
$W$ between the ideal leads and the extended molecule
switched on.  The scattering amplitudes $t_{jk}$ from
the ideal source lead $k$ to drain lead $j$ at energy
$E$ were extracted from the scattering eigenstates
$|\Psi_{k}\rangle$ and the transmission probabilities
that enter eq. (\ref{Landauer}) were then calculated
from
\begin{equation}
T_{x}(E,V) = \sum_k \sum_j \left | \frac{v_j}{v_k} \right |
|t_{jk}|^2
\label{eq:multiTx}
\end{equation}
where ${v_k}$ and ${v_j}$ are the electron group velocities
in ideal leads $k$ and $j$ respectively at energy
$E$. The sum over $j$ in Eq. (\ref{eq:multiTx}) is
over those drain leads connected to electrode $x$,
i.e., the gold metal film or the Si substrate as
discussed above, while that over $k$ is over the
leads  connected to the STM tip.

\section{Results}
\label{Results}

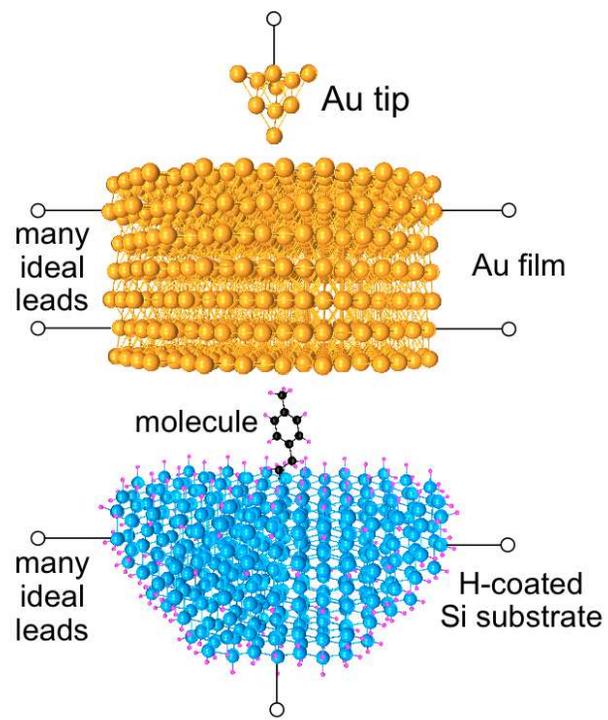
\begin{figure*}
\caption{(Color on line). The extended molecule
studied in the BEEM transport calculations for the
case where the molecule is
ethylmethylbenzene.\cite{MacMolPlt} The STM tip is
represented by 10 Au atoms in a (111) geometry and
terminates in a single atom. The Au film is
represented by a cylindrical cluster of 591 Au atoms
in a bulk crystal geometry that presents
(111)-oriented facets to the STM tip and the
molecule. The substrate is represented by an
approximately hemispherical crystallite of 390 Si
atoms. The Si dangling bonds are passivated with H
atoms and a flat (111) surface faces the molecule.
The molecule bonds to the Si substrate through a
single C-Si covalent bond. A large number of single
channel ideal leads (most of them not shown)
representing the source and drain electrodes are
attached to the atomic valence orbitals of
appropriate surface atoms of the clusters 
representing the STM tip, Au film, silicon
substrate and passivating H atoms.}
\label{Fig_1}
\end{figure*}

For the terminal atom of the gold STM tip over an
atom at the center of the top surface of the Au
cluster that represents the gold film in
Fig.\ref{Fig_1}, and separated from that surface
atom by a nearest neighbor distance of gold
(2.88\AA), the STM tip and gold film are connected by
an atomic quantum point contact. For this geometry
the present calculations of the electron transmission
probability  $T_{\mathrm{FILM}}(E,0)$ between the
tip and gold film at zero applied bias yield values
close to 1 (per spin) for energies $E$ around the
Fermi level of gold, consistent with previous 
experimental\cite{AuatomQPCexpt,AuatomQPCexptandtheo}
and
theoretical\cite{AuatomQPCexptandtheo,AuatomQPCtheory}
studies of gold atomic point contacts. However, in
BEEM experiments the STM tip is normally further from the gold
film, at distances in the STM tunneling regime. Unless stated
otherwise, the results of the BEEM calculations to be presented
here will be for such a larger separation (4.33\AA)
between the tip atom and the gold film for which the
calculated the electron transmission probability 
$T_{\mathrm{FILM}}(E_F,0)$ between the tip and gold
film at the gold Fermi energy and zero applied bias
is approximately
$10^{-2}$.  The calculated electron transmission
probabilities $T_{\mathrm{BEEM}}(E,V)$ from the
electron source to the drain via the STM tip, gold
film, molecule  and silicon substrate are shown in
Fig. \ref{Fig_2} for two representative molecules
that are shown in the insets. The solid (dotted)
curves are for zero  (2V) bias $V$ between the STM
tip and gold film. The Fermi level of the gold film
is at E=0 in all cases, and the silicon conduction
band edge at the silicon surface $E_C$ is indicated
by the vertical dashed line. The lower plot is for
decanethiolate
($\mathrm{C}_{10}\mathrm{H}_{20}\mathrm{S}$) thiol
bonded to the gold film below a hollow site between 3
gold atoms and attached to the silicon substrate by a
single carbon-silicon bond. This molecule is
insulating, having a large HOMO-LUMO gap, and thus
its transmission in a BEEM experiment is predicted to
be very small resulting in $T_{\mathrm{BEEM}}(E,V)
\le 10^{-11}$ in the range shown. The upper plot is
for another chain molecule 
($\mathrm{C}_{10}\mathrm{H}_{18}\mathrm{S}$) also
with 10 carbon atoms and bonded similarly to the gold
film and Si substrate, but in this case the 5th and
6th carbon atoms each have one H atom attached
instead of two so that there is a double bond
connecting these two C atoms; see upper inset. This
double C-C bond gives rise to a molecular state near
$E=1.65$eV where the transmission
$T_{\mathrm{BEEM}}(E,V)$ is greatly enhanced (at its
peak by almost  seven orders of magnitude relative to
that for the decanethiolate) although it is still
{\em very} weak ($<10^{-5}$) in absolute terms, as expected
for a transmitting state associated with a double bond surrounded by
large  potential barriers. However, there is no other 
resonantly transmitting state in this system in the
energy range above and reasonably close to the Fermi level
of the gold film, the range relevant to BEEM 
experiments. Therefore despite its weakness this resonance
is the dominant BEEM transport mechanism for this system 
as will be seen below. Notice
that the application  of a 2V bias does not change
the transmission probabilities
$T_{\mathrm{BEEM}}(E,V)$ by much on the scale of
Fig. \ref{Fig_2} (less than a factor of 2) because
the main transport bottleneck here is the molecule
and the application of a bias between the tip and
gold film does not affect the molecule significantly.

In BEEM experiments the distance between the tip and
gold film is not known accurately, as is the case for
most STM experiments where only {\em changes} in the
tip-surface separation are measured. However, the
transmission  probabilities $T_{\mathrm{BEEM}}$ and
$T_{\mathrm{FILM}}$ and currents
$I_{\mathrm{BEEM}}$ and $I_{\mathrm{FILM}}$ depend
strongly on the tip-surface separation. This makes a
direct comparison between calculated and experimental
values of these quantities problematic. However, it
is intuitively plausible that 
$I_{\mathrm{BEEM}}$ and $I_{\mathrm{FILM}}$ should
scale in the same way with the tip-surface separation
in the tunneling regime and thus
$I_{\mathrm{BEEM}}/I_{\mathrm{FILM}}$ should be
nearly independent of the tip-surface separation.
This has been confirmed by calculations carried out
in the present study. Thus the ratio
$I_{\mathrm{BEEM}}/I_{\mathrm{FILM}}$ lends  itself
better than $I_{\mathrm{BEEM}}$ or
$I_{\mathrm{FILM}}$ to comparison between theory and
experiment and it will therefore be the focus of
attention in what follows. Another advantage of
studying this quantity theoretically is that
spectroscopic BEEM (i.e. BEES) measurements of
$I_{\mathrm{BEEM}}$ vs. the bias voltage applied
between the STM tip and metal film are normally taken
not at constant tip-surface separation but at
constant tip current $I_{\mathrm{tip}}$, which, for
the systems studied here, differs negligibly from the
film current $I_{\mathrm{FILM}}$. Therefore because
$I_{\mathrm{BEEM}}/I_{\mathrm{FILM}}$ is
approximately independent of the tip-surface
separation, plots of
$I_{\mathrm{BEEM}}/I_{\mathrm{FILM}}$ vs. bias
voltage should differ from experimental plots of
$I_{\mathrm{BEEM}}$ vs. bias by only a scale factor
that is independent of the bias, making qualitative
comparisons between calculations and BEES
current-voltage characteristics appropriate. A
typical experimental BEEM setup can readily detect
BEEM currents with current ratios
$I_{\mathrm{BEEM}}/I_{\mathrm{tip}}$ greater than
roughly $10^{-5}$; this number will be referred to
below as the ``nominal BEEM sensitivity threshold"
(NBT).

In the present calculations the silicon cluster that
represents the silicon substrate in the extended
molecule is of necessity small even though it
includes hundreds of atoms; the silicon hemisphere
shown in Fig.\ref{Fig_1} has a radius  slightly less
than 1.5nm. Because of its small size it transmits
electrons appreciably by quantum tunneling even at
energies below the conduction band edge
$E_C$ in Fig.\ref{Fig_2} where there are no
eigenstates of the extended molecule with a strong
silicon content. Because of this the calculated
transmission $T_{\mathrm{BEEM}}(E,V)$ shown in in
Fig.\ref{Fig_2}, although it is extremely weak below
$E_C$, does not decline all the way to zero for $E <
E_C$. Thus while the present model should describe
the BEEM electron transmission reasonably well at
energies higher than $E_C$, it does not describe the
details of the $onset$ of BEEM current at the
threshold voltage $V = E_C$ accurately for
macroscopic silicon substrates. However, as will be
seen below, for the molecules studied here, the
current ratios $I_{\mathrm{BEEM}}/I_{\mathrm{tip}}$
in the near-threshold regime are very far below
nominal BEEM sensitivity threshold, and detailed
study of this regime is therefore left for future
work. Here, for simplicity, in calculating
$I_{\mathrm{BEEM}}$ from Eq. (\ref{Landauer}),
$T_{\mathrm{BEEM}}(E,V)$ will be set to zero for $E <
E_C$.         
  
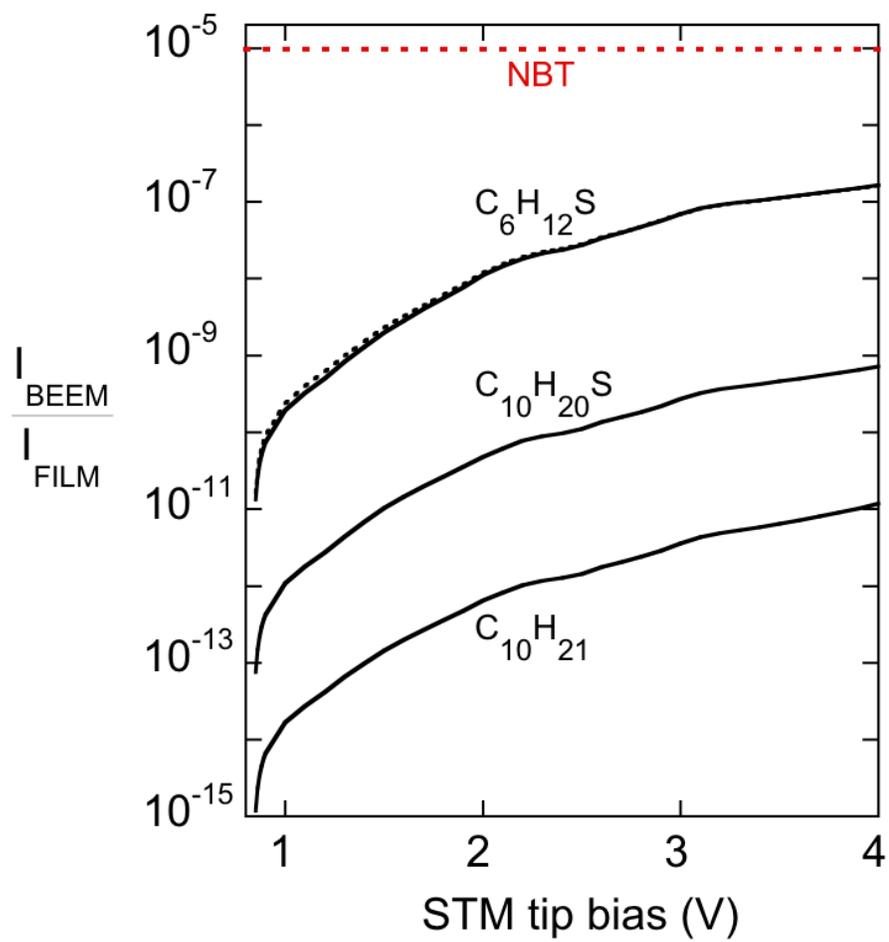
\begin{figure*}
\caption{(Color on line).
Calculated BEEM to film current ratios vs.
tip bias voltage $V$ for some saturated molecules:
Decane ($\mathrm{C}_{10}\mathrm{H}_{21}$),
decanethiolate
($\mathrm{C}_{10}\mathrm{H}_{20}\mathrm{S}$) and
hexanethiolate
($\mathrm{C}_{6}\mathrm{H}_{12}\mathrm{S}$). The red
horizontal dashed line is the nominal BEEM
sensitivity threshold. Solid curves are 
for a separation of 4.33{\AA} between the STM tip atom and 
the gold film. The (black) dotted curve for
$\mathrm{C}_{6}\mathrm{H}_{12}\mathrm{S}$
is for a 5.05{\AA}
separation.}
\label{Fig_3}
\end{figure*}

The calculated current ratios
$I_{\mathrm{BEEM}}/I_{\mathrm{tip}}$ for some
representative saturated chain molecules are shown
in Fig. \ref{Fig_3}.
$\mathrm{C}_{10}\mathrm{H}_{21}$ is decane bonded to
the silicon substrate through a single C-Si bond and
not bonded chemically to the gold film. Even at an
STM tip bias of 4V the calculated
$I_{\mathrm{BEEM}}/I_{\mathrm{tip}}$ for this
molecule is 6 orders of magnitude below the nominal
BEEM sensitivity threshold. For
$\mathrm{C}_{10}\mathrm{H}_{20}\mathrm{S}$
(decanethiolate depicted in the lower inset of Fig.
\ref{Fig_2}) the thiol bond to the Au film results in
predicted BEEM currents two orders of magnitude
larger than for $\mathrm{C}_{10}\mathrm{H}_{21}$.
However, the calculated current ratios
$I_{\mathrm{BEEM}}/I_{\mathrm{tip}}$ for both
$\mathrm{C}_{10}\mathrm{H}_{21}$ and
$\mathrm{C}_{10}\mathrm{H}_{20}\mathrm{S}$ are still
so low as to be well below the NBT not only for
single molecules but even for self-assembled
molecular monolayers where hundreds of molecules may
be contributing together to the BEEM current. For
$\mathrm{C}_{6}\mathrm{H}_{12}\mathrm{S}$
(hexanethiolate which is similar to decanethiolate
but with fewer carbon atoms and so presents a thinner
tunnel barrier than decanethiolate) the calculated
current ratios $I_{\mathrm{BEEM}}/I_{\mathrm{tip}}$
are two orders of magnitude higher than for
decanethiolate but still well below the NBT. Thus
detecting a BEEM signal from single alkane or
alkanethiolate molecules on silicon or even from
self-assembled monolayers of these molecules is
expected to be challenging experimentally except for
quite short molecules of these types.  This is quite
reasonable given the well known insulating
nature of alkanes and alkanethiols.
  
While the solid curves in Fig. \ref{Fig_3}  are 
for a separation of 4.33{\AA} between the STM tip atom and 
the gold film, the (black) dotted curve for
$\mathrm{C}_{6}\mathrm{H}_{12}\mathrm{S}$ is for
a 5.05{\AA}  separation at which the calculated 
transmission $T_{\mathrm{FILM}}$ between
the STM tip and gold film is lower by a factor
of approximately 35. Notice that the calculated values
of $I_{\mathrm{BEEM}}/I_{\mathrm{tip}}$ are almost
the same for the two separations, illustrating the
insensitivity of the ratio $I_{\mathrm{BEEM}}/I_{\mathrm{tip}}$
to the separation between the STM tip and gold
film in the tunneling regime, as has been discussed above. 

The calculated current ratios
$I_{\mathrm{BEEM}}/I_{\mathrm{tip}}$ for some
representative unsaturated molecules are shown in
Fig. \ref{Fig_4}. Transmission resonances due to
molecular states associated with the unsaturated C-C
bonds in these molecules give rise to dramatic
increases in the BEEM current at bias voltages
(indicated by arrows) at which the Fermi level of the
STM tip crosses the resonant molecular energy levels.
As a result, even for
$\mathrm{C}_{10}\mathrm{H}_{18}\mathrm{S}$ (depicted
in the upper inset of Fig. \ref{Fig_2}) the largest
value of $I_{\mathrm{BEEM}}/I_{\mathrm{tip}}$ is
predicted to be within an order of magnitude of the
NBT. For $\mathrm{C}_{6}\mathrm{H}_{10}\mathrm{S}$
(similar to
$\mathrm{C}_{10}\mathrm{H}_{18}\mathrm{S}$ but with
a shorter C chain and only one H atom bonded to C
atoms 3 and 4) and for
$\mathrm{C}_{9}\mathrm{H}_{11}$ (ethylmethylbenzene
depicted in Fig. \ref{Fig_1})
$I_{\mathrm{BEEM}}/I_{\mathrm{tip}}$ is predicted to
exceed the NBT for bias voltages above the threshold
for resonant transport.\cite{LUMO} 
Thus experimental observation
of single buried molecules such as these in BEEM
experiments should be feasible with conventional BEEM
equipment. Because the predicted current ratios
$I_{\mathrm{BEEM}}/I_{\mathrm{tip}}$ for saturated
molecules (Fig. \ref{Fig_3}) are much smaller than
those for unsaturated molecules of similar length
(Fig. \ref{Fig_4}) it should be possible to study
individual buried unsaturated molecules by BEEM and
BEES by including a few unsaturated molecules in a
self-assembled monolayer of saturated molecules of
similar size on silicon under a thin gold film, the
saturated molecules behaving as an insulating
background surrounding the unsaturated conducting
molecules. I.e., BEEM and BEES experiments extending
to {\em buried} single molecules the classic STM
studies\cite{Bumm,Donhauser} of the properties of
single unsaturated molecules embedded in
self-assembled monolayers of saturated molecules on
gold substrates should be feasible. However, in
interpreting such experiments it is important to
note that the calculated current ratios
$I_{\mathrm{BEEM}}/I_{\mathrm{tip}}$ shown in Fig.
\ref{Fig_4} for unsaturated molecules rise with
increasing bias from values far below the NBT.
Because of this, experimentally observed values of
the threshold voltage for the onset of the BEEM
current are likely to depend on the sensitivity of
the BEEM equipment unless equipment capable of
detecting extremely weak BEEM currents is used.

As in Fig. \ref{Fig_3}, the solid curves in Fig. \ref{Fig_4} 
are for a separation of 4.33{\AA} between the STM tip atom and 
the gold film while the (black) dotted curve for
$\mathrm{C}_{6}\mathrm{H}_{10}\mathrm{S}$ is for
a 5.05{\AA} separation at which the calculated 
transmission $T_{\mathrm{FILM}}$ is again lower by a factor
of approximately 35. Once again the calculated values
of $I_{\mathrm{BEEM}}/I_{\mathrm{tip}}$ are almost
the same for the two separations, illustrating the
insensitivity of the ratio $I_{\mathrm{BEEM}}/I_{\mathrm{tip}}$
to the separation between the STM tip and gold
film in the tunneling regime, this time for a resonantly transmitting
molecule.

In the course of the present work, preliminary
results have also been obtained for some defects
that may occur in molecular SAMS on silicon covered
with gold: It was found that small interstitial gold
clusters embedded between decanethiolate molecules
of the SAM and also single gold atoms substituting
for H atoms of decanethiolate molecules should give
rise to BEEM transmission resonances and associated
features in $I_{\mathrm{BEEM}}/I_{\mathrm{tip}}$
qualitatively similar to shown in Fig. \ref{Fig_4}.
However, these resonant features in
$T_{\mathrm{BEEM}}$ and $I_{\mathrm{BEEM}}$ were
found to be much weaker for the defects that were
studied (a 6 atom Au cluster roughly equidistant from
the Au film and Si substrate and a single gold atom
substituting for a H atom on the 5th C atom [from the
Si substrate] of decanethiolate) than those of the
unsaturated 
$\mathrm{C}_{10}\mathrm{H}_{18}\mathrm{S}$ molecule.

Finally, it is interesting that the recent
experimental studies\cite{KavanaghGaAs} of
gold-covered octanedithiol molecules on GaAs
substrates observed BEEM current-voltage
characteristics qualitatively similar to the resonant
behavior in Fig.
\ref{Fig_4}. However, since the experiments were on
saturated molecules and the BEEM signal was observed
only in a few very small patches of the sample the
origin of the observed behavior is unclear.

\section{Conclusions}
\label{conclusions}

A better understanding of the
electronic and structural properties of buried molecular  
layers and
single molecules is essential for the development of
molecular nanoelectronics.  In this paper the
feasibility of imaging single buried molecules and
measuring their electronic spectra has been explored
theoretically. It was found that the insulating
nature of single alkane and alkanethiolate molecules
chemisorbed on silicon substrates and covered with
gold should make them difficult to detect by
ballistic electron emission miscroscopy and
spectroscopy for all but very short molecules of this
kind. However BEEM currents resonantly transmitted
through molecules containing double C-C bonds or
aromatic rings should be large enough to be
accessible to present day BEEM  equipment, and
predictions of the BEEM current-voltage
characteristics for some examples of molecules
of each type have been presented.
The predicted BEEM current-voltage characteristics for unsaturated 
molecules exhibit observable features
(marked by the arrows in Fig. \ref{Fig_4}) due to resonantly
transmitting electronic states of the extended molecule that lie above
the Fermi level of the metal overlayer. The energy differences between
these resonant electronic states and the Fermi level of the metal film
are given by $|eV|$ where $V$ is the bias voltage between
the STM tip and metal film at which such a feature occurs. Thus
BEEM current-voltage characteristics can be used to measure
the electronic spectra of buried unsaturated molecules in
the environment between the metal and semiconductor layers.
Since no bias voltage is applied between the metal overlayer and
semiconductor substrate that surround the molecule these BEEM
spectra differ from those obtained using conventional diode
measurements in which the bias voltage is applied {\em across the
molecule} and therefore modifies the electronic structure that is being
measured.

At the present time rigorous
theoretical techniques for molecular electronics, and
especially molecular
transport calculations, have yet to be developed. Thus the present
study relies on a combination of semi-empirical and {\em ab-initio}
methods as well as intuitively reasonable approximations and
assumptions. The results presented should facilitate the design and
interpretation of experiments by providing theoretical answers to
some rather basic questions regarding BEEM and BEES of molecules that
have not been addressed previously.
Further theoretical work on this topic as well as experiments testing
the present predictions are clearly desirable, and it is hoped that the
present exploration will stimulate such efforts.

\section*{Acknowledgments}

I thank K. Kavanagh and H. Dalgleish for discussions.
This research was supported by
the Canadian Institute for Advanced Research and NSERC.

%%%%%%%%%%%%%%%%%%%%%%%%%%%%%%%%%%%%%%%
%
%

%
% Fig. 1extended molecule
%

%Fig. 2

\begin{figure*}
\caption{(Color on line). Calculated electron
transmission probabilities
$T_{\mathrm{BEEM}}(E,V)$ from electron source  to
drain via STM tip, gold film, molecule and and
silicon substrate vs. electron energy $E$. Solid
(dotted) curves are for at zero (2V) bias $V$
between STM tip and gold film. The gold film Fermi
energy is at
$E=0$ and the Si conduction band edge at the Si
surface $E_C$ is shown by the dashed vertical line.
The lower plots are for the decane thiolate
$\mathrm{C}_{10}\mathrm{H}_{20}\mathrm{S}$ molecule
in the lower inset. The upper plots are for the
$\mathrm{C}_{10}\mathrm{H}_{18}\mathrm{S}$ molecule
in the upper inset that exhibits resonant
transmission due to the presence of a double C$-$C
bond.}
\label{Fig_2}
\end{figure*}

%
%
% Fig. 3
%

%
%
% Fig. 4
%
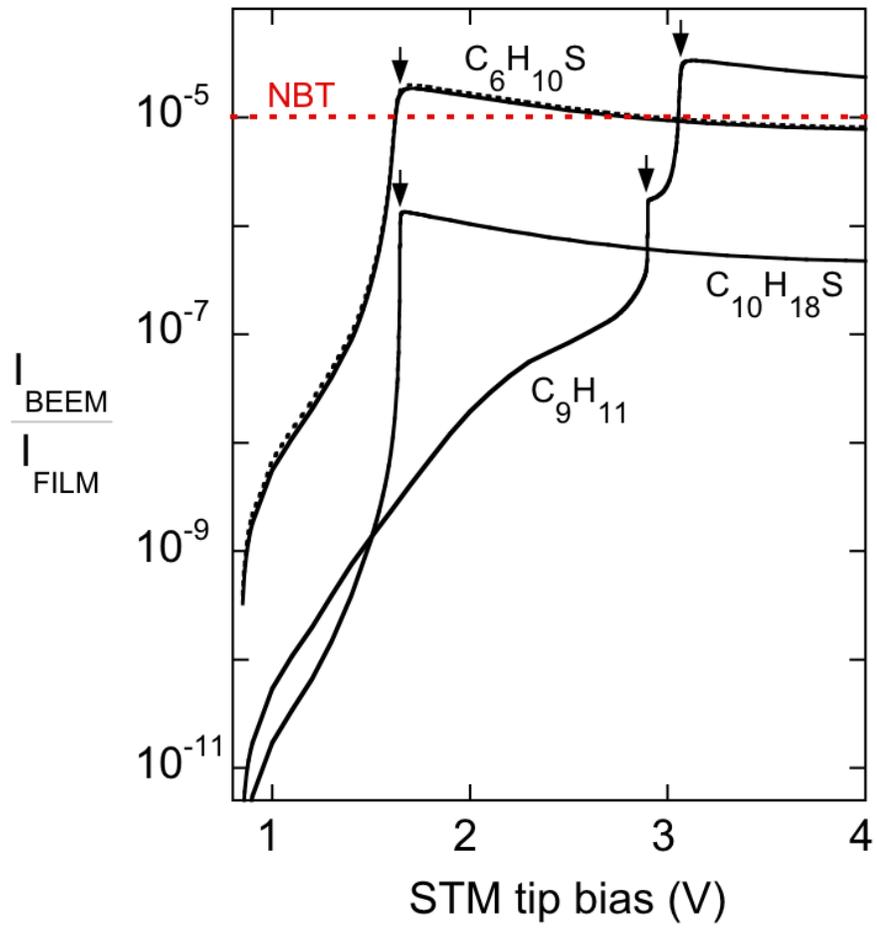
\begin{figure*}
\caption{(Color on line).
Calculated BEEM to film current ratios vs.
tip bias voltage $V$ for some unsaturated chain
molecules
$\mathrm{C}_{10}\mathrm{H}_{18}\mathrm{S}$ (shown
in the upper inset of Fig. \ref{Fig_2}) and 
 $\mathrm{C}_{6}\mathrm{H}_{10}\mathrm{S}$, and for
ethylmethyl benzene
$\mathrm{C}_{9}\mathrm{H}_{11}$ shown in Fig.
\ref{Fig_1}. Arrows indicate values of the bias
voltage at which the STM tip Fermi level crosses the
energies of molecular resonant states.  The red
horizontal dashed line is the nominal BEEM
sensitivity threshold. Solid curves are 
for a separation of 4.33{\AA} between the STM tip atom and 
the gold film. The (black) dotted curve for
$\mathrm{C}_{6}\mathrm{H}_{10}\mathrm{S}$
is for a 5.05{\AA}
separation.}
\label{Fig_4}
\end{figure*}


\begin{thebibliography}{100}
\bibitem{meV}In BEEM experiments there may be a
potential difference of a few meV between the
semiconductor substrate and the metal overlayer,
however, this potential difference is normally much
smaller than the potential difference between the STM
tip and metal overlayer. Thus its effect on the BEEM
current is usually very small and will be neglected
in the present theoretical study.
\bibitem{temp}The temperature $T$ in the Fermi
functions was set to 0 in the calculations reported
here.
\bibitem{LUMO}The two arrows associated with
the BEEM current-voltage characteristic of $\mathrm{C}_{9}\mathrm{H}_{11}$
in Fig. \ref{Fig_4} mark features that
are due to transmission resonances associated with the lowest
unoccupied molecular orbital (LUMO) and the LUMO+1 orbital of
the molecule.
\end{thebibliography}
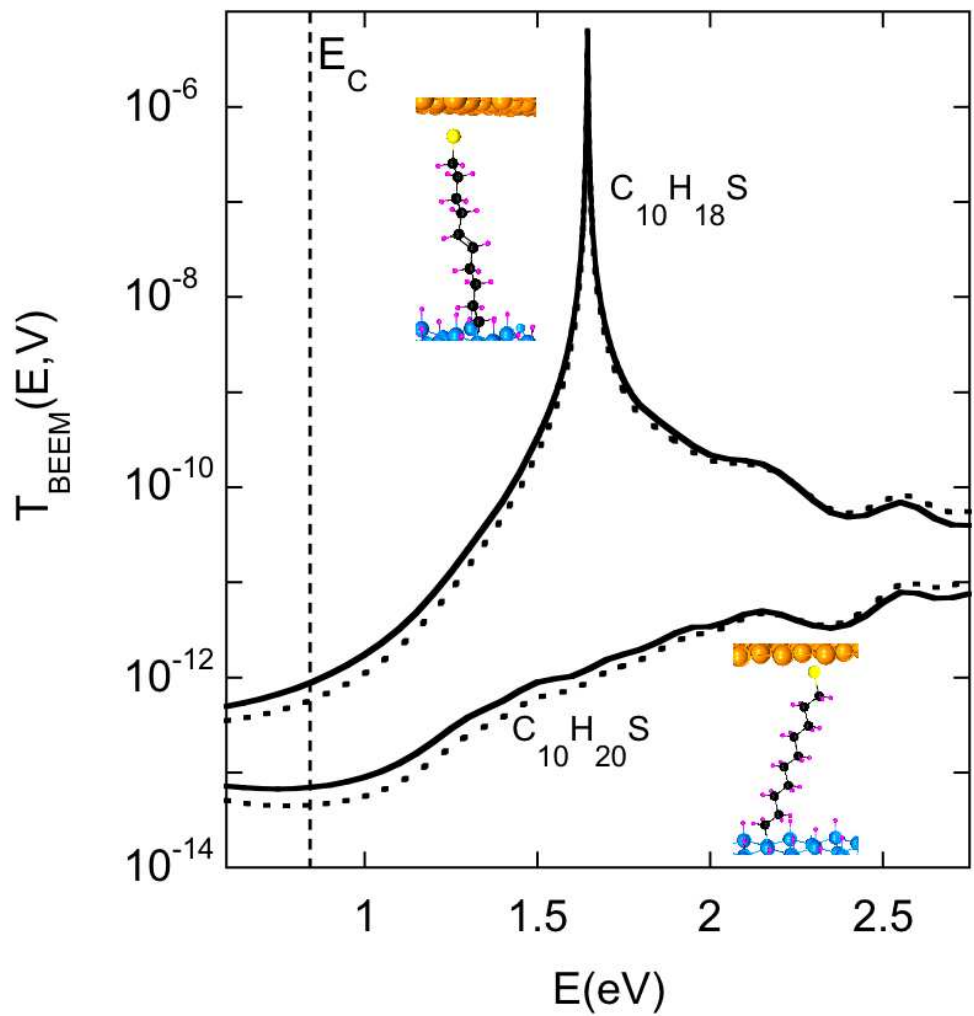
\end{document}